\documentclass[journal]{IEEEtran}
\usepackage{cite}
\usepackage{amsmath,amssymb,amsfonts}
\usepackage{algorithmic}
\usepackage{graphicx}
\usepackage{textcomp}
\usepackage{comment}
\usepackage{xcolor}
\usepackage{url}
\usepackage[caption = false]{subfig}
\usepackage{footnote}
\usepackage{makecell}
\usepackage{siunitx}
\usepackage{amsfonts} 
\DeclareMathSymbol{\shortminus}{\mathbin}{AMSa}{"39}

\makesavenoteenv{tabular}
\makesavenoteenv{table}
\def\BibTeX{{\rm B\kern-.05em{\sc i\kern-.025em b}\kern-.08em
    T\kern-.1667em\lower.7ex\hbox{E}\kern-.125emX}}
\begin{document}
\raggedbottom
\title{An Asynchronous and Low-Power True Random Number Generator using STT-MTJ}

\author{Ben~Perach and Shahar~Kvatinsky\vspace{-0.8cm}
\thanks{The authors are with the Andrew and Erna Finci Viterbi Faculty of Electrical Engineering, Technion - Israel Institute of Technology, Haifa, Israel 3200003. e-mail: benperach@campus.technion.ac.il, shahar@ee.technion.ac.il.
This research is partially supported by the ERC under the European Union's Horizon 2020 Research and Innovation Programme (grant agreement no. 757259), by the Technion Hiroshi Fujiwara Cyber Security Research Center, and by the Israel Cyber Bureau.}
}

\markboth{IEEE Transaction on Very Large Scale Integration Systems,~Vol.~XX, No.~XX, XXXX-XX}
{Perach \MakeLowercase{\textit{et al.}}: An Asynchronous and Low-Power True Random Number Generator using STT-MTJ}

\maketitle

\begin{abstract}
The emerging Spin Transfer Torque Magnetic Tunnel Junction (STT-MTJ) technology exhibits interesting stochastic behavior combined with small area and low operation energy. It is, therefore, a promising technology for security applications, specifically the generation of random numbers. In this paper, STT-MTJ is used to construct an asynchronous true random number generator (TRNG) with low power and a high entropy rate. The asynchronous design enables decoupling of the random number generation from the system clock, allowing it to be embedded in low-power devices. The proposed TRNG is evaluated by a numerical simulation, using the Landau\textendash Lifshitz\textendash Gilbert (LLG) equation as the model of the STT-MTJ devices. Design considerations, attack analysis, and process variation are discussed and evaluated. We show that our design is robust to process variation, achieving an entropy generating rate between 99.7Mbps and 127.8Mbps with 6-7.7 pJ per bit for $90\%$ of the instances. 
\end{abstract}


\IEEEpeerreviewmaketitle

\section{Introduction}
\label{sec:introduction}
\IEEEPARstart{S}{ecurity} is a major concern in modern digital systems. One of the main tools used in security is cryptography, used to encode information that only authorized entities can access. However, security applications need to be implemented with caution. Changing the cryptographic algorithm or its assumptions, even to a limited extent, can compromise the entire system. One such crucial part of cryptographic algorithms is the generation of the cryptographic keys~\cite{RFC4086,Menezes,Goldberg1996,Gutterman2006,Kelsey1998}.

The key of a cryptographic algorithm is the secret of the encryption scheme. The algorithm itself is assumed to be publicly known, and the key is the only missing information needed to reveal the encrypted data~\cite{Menezes}. Hence, an adversary will try to obtain the key. Since the key is of a finite size, the number of possible values for the key is finite as well, and if this number is too small, an adversary can try them all. Additionally, if the adversary has partial knowledge of the key, such as some mathematical conditions between the key bits, this information can be used to reduce the number of options~\cite{Goldberg1996,Kelsey1998,Gutterman2006}. Hence, it is desirable to generate a random key with a uniform distribution on all of its possibilities, so an adversary will have to try all of the options without a defined order. Processes that can generate a random number as the key are called \textit{Random Number Generators} (RNGs). Note that the design of the RNG itself, as part of the encryption scheme, is also assumed to be publicly known. 

One type of RNG is a TRNG, a \textit{true} random number generator~\cite{Koc2008,BHUNIA2019}. A TRNG is based on a physically random process (\textit{e.g.}, thermal noise), and the TRNG extracts that randomness to a usable form, such as digital numbers. The TRNG approach for generating random numbers is attractive since the generated number cannot be inferred from the state of the system but can only be predicted from the distribution of the physical random process~\cite{RFC4086,Menezes}.

Current TNRGs use CMOS logic, such as ring oscillators~\cite{KYang} or metastable latches~\cite{IntelTRNG}, as their source for randomness. However, emerging technologies~\cite{Akinaga2010,Wong2010,Wong2012,Wang2018,Hu2015} offer new and interesting alternatives, due to their smaller area and lower power consumption when compared to transistors. A small area and low power TRNG, based on a random process in emerging technologies, will reduce power consumption or enable secure communication for small or low-power electronic devices (\textit{e.g.}, Internet-of-Things devices and mobile devices). One such technology is the \textit{Spin Transfer Torque Magnetic Tunnel Junction} (STT-MTJ)~\cite{Akinaga2010,Wang2018,Hu2015}. As an emerging memory technology, STT-MTJ (or STT-MRAM) has relatively low operating energy and small area, and its switching time stochasticity has been thoroughly studied~\cite{Devolder}.

While previously proposed STT-MTJ based TRNGs~\cite{Vatajelu,Fukushima,Oosawa,Qu2017,Wang2016} require a strict time measurement to achieve high randomness, we propose an \textit{asynchronous} TRNG. The proposed design relies on discharging a capacitor simultaneously through several STT-MTJ devices. The process ends when the capacitor is sufficiently discharged; the generated random number is extracted from the final state of the STT-MTJ devices.
Since the capacitor is discharged asynchronously, the random number generation is independent of a clock signal.
The only time measurement required in the TRNG design is the duration to discharge the capacitor, which can be approximated by defining a minimum time.
Waiting for more than the minimum time does not influence the randomness of the output.
Therefore, this design can be embedded in low-frequency (\textit{i.e.}, low-power) devices without loss of randomness. Furthermore, the use of several STT-MTJ devices increases the extracted randomness in each operation and improves the robustness to process variation.   

To evaluate the proposed design, numerical simulations of the stochastic physical model of the STT-MTJ were performed. The randomness of the proposed design and the effect of internal and external influences were measured, including the robustness of the design to process variation. Possible attack venues are discussed and mitigation options are presented. To the best of our knowledge, this is the first study of an STT-MTJ based TRNG that includes analysis of attacks, essential for every security-related work\footnote{Attack analysis was previously conducted only on other STT-MTJ based security devices, such as physically unclonable functions~\cite{Ghosh}.}.

\section{Background}
\label{sec:background}
\subsection{True Random Number Generators}
\label{subsec:trng}

RNGs are divided into two main groups, \textit{Pseudo-Random Number Generators} (PRNGs) and \textit {True Random Number Generators} (TRNGs). PRNGs are \textit{deterministic} algorithms that only appear to generate a random sequence of numbers.
A cryptographic key generated by a PRNG might compromise the encryption since the PRNG outputs are inherently connected~\cite{Goldberg1996,Kelsey1998,Gutterman2006}, although some PRNGs are considered to be sufficiently secure for cryptographic use~\cite{NIST90A,Kan2007}. TRNGs are designed to extract a random behavior of some physically random process~\cite{KYang,IntelTRNG,Koc2008,BHUNIA2019}, resulting in true randomness that can be explained according to some physical laws. The output of a TRNG can only be predicted according to the physical process probability distribution, even if all the information about the system (register values, voltage levels, \textit{etc.}) is known prior to the TRNG operation.

CMOS TRNGs often use ring oscillators~\cite{KYang,Yingjie} or metastable latches\cite{IntelTRNG} to generate random numbers. A ring oscillator (RO) is a chain with an odd number of NOT gates, where the output of the last NOT gate is the input of the first NOT gate, resulting in a ring of gates. Since the number of NOT gates is odd, all the outputs of the NOT gates oscillate between logic high and logic low. However, due to noise in the transistors, the rise and fall times of the gates are randomly changed, resulting in frequency variation of the oscillation. RO based TRNGs use this frequency variation as the source of randomness: for example, they might compare several ROs~\cite{Yingjie} or measure the time until the occurrence of an RO-related event~\cite{KYang}. TRNGs based on metastable latches force a latch into an unstable equilibrium state and then release it. The stable state the latch will end in depends on random noise and therefore a random number is generated.   

The quality of the randomness of a TRNG can be measured by its output properties, which are ideally independent and uniformly distributed. 
In practice, this need not be the case for cryptographic use, since the 
dependencies and nonuniform distribution can be compensated for by post-processing the output. Such post-processing methods are referred to as \textit{randomness extractors}~\cite{Salil,Kwok2011}.
However, the closer to independent and uniformly distributed the TRNG output is, the simpler the extractor can be. Even if the output is uniform under regular operation, randomness extractors are still commonly used to compensate for real-world effects (\textit{e.g.}, process variations, wear-out, and interference) that might reduce the randomness of the TRNG output~\cite{Kwok2011,BHUNIA2019}.

Other important properties of a TRNG include robustness to process and environmental variations and a high generating rate. An adversary might change environmental parameters (\textit{e.g.}, electromagnetic field, temperature) to interfere with the operation of the TRNG and reduce its randomness. To guarantee that a TRNG is secure, it is critical to identify the underlying random physical process and the factors affecting it and test the TRNG under these factors. Additionally, to ensure correct operation, statistical tests, referred to as \textit{online tests}, are often performed on the TRNG output during run-time, which means that they need to be lightly implemented, making them less thorough than the statistical tests performed at design time.

New emerging technologies, such as the STT-MTJ~\cite{Akinaga2010,Wang2018,Hu2015}, show small area and low operation energy compared to transistors. Additionally, they exhibit ample stochasticity in their operation~\cite{Devolder}, making them interesting candidates as the randomness source for new TRNG designs.

\subsection{STT-MTJ Devices}
\label{subsec:stt-mtj}

\begin{figure}[tb]
\centering
\hspace{-1.2cm}
\includegraphics[width = 8.2cm,height=4.0cm]{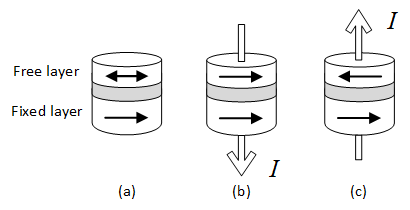}
\caption{In-plane STT-MTJ device and its operation. (a) An STT-MTJ device structure with the free layer (upper), the tunnel barrier layer (middle), and the fixed layer (bottom). (b) The P and (c) AP states and the associated currents to switch toward the state.}
\label{pic:MTJ}
\vspace{-0.4cm}
\end{figure}

A spin transfer torque (STT) magnetic tunnel junction (MTJ) is a device composed of two ferromagnetic layers with a tunnel barrier layer between them~\cite{Devolder,Vincent}. One ferromagnetic layer, the \textit{fixed layer}, has a fixed magnetization direction. The other ferromagnetic layer, the \textit{free layer}, can switch its magnetization direction. In this paper, in-plane MTJs are used, where the magnetization direction of the ferromagnetic layers is in the plane of the layers. Figure~\ref{pic:MTJ}(a) illustrates the MTJ structure. The direction of the free layer magnetization can be changed by a current through the device, and it has two stable states, parallel (P, Figure~\ref{pic:MTJ}(b)) or anti-parallel (AP, Figure~\ref{pic:MTJ}(c)) to the direction of the fixed layer. The direction of the current determines the change in the magnetization direction. Other states (\textit{i.e.}, other directions of the free layer magnetization) are unstable.

The spin-transfer torque mechanism enables the switching of orientation of the free layer magnetization. The electrons passing through a ferromagnetic layer tend to align their magnetic moment in the direction of the magnetization of the layer.
Thus, electrons that pass through the fixed layer first are aligned with its magnetization direction. When these electrons reach the free layer, its magnetization direction shifts towards the P state due to magnetic moment conservation (Figure~\ref{pic:MTJ}(b)). In the other current direction, electrons are reflected with magnetic moment direction opposite to that of the fixed layer and change the free layer to the AP state (Figure~\ref{pic:MTJ}(c)). However, a damping process pulls the free layer magnetization to the closest stable state, requiring a sufficiently strong current for adequate time to enable a switch between the stable states. 

The switching process between the P and AP states is random~\cite{Devolder} due to the thermal fluctuations in the ferromagnetic layers. Although the current through the MTJ pushes the magnetization of the free layer to a certain stable state (through unstable intermediate states), thermal fluctuations will make the path to that state random, resulting in a random switching time. Even if no current is applied, the state of the STT-MTJ fluctuates constantly since the thermal fluctuations occur regardless of the existence of the current. 

The state of the MTJ also determines its resistance, where the P state resistance is marked as $R_{on}$, the AP state is marked as $R_{off}$, and $R_{on} < R_{off}$. The resistance of the MTJ, when it is in a state other than P or AP, is between $R_{on}$ and $R_{off}$ and its exact value depends on the state~\cite{Slonczewski1989}. To determine the state of the MTJ, a low voltage can be applied across it (sufficiently low not to incur a switch), the current can be measured, the resistance of the MTJ can be extracted (by Ohm's law), and the state of the MTJ can be inferred. 

To model the operation of the entire STT-MTJ, the magnetization of the free layer is usually approximated to a single domain. The phenomenological Landau-Lifshitz-Gilbert (LLG) equation~\cite{Gilbert}, with the addition of a stochastic term for the thermal fluctuations~\cite{GarcuaPalacios} and Slonczewski's STT term~\cite{Slonczewski1996}, can accurately describe the dynamics of the magnetization of the free layer.
For current pulses with low or high current magnitudes, approximations and models for the distribution of the switching time exist~\cite{Apalkov,Li,Sun}. For current pulses with intermediate current magnitudes, approximations for the switching time distribution and other models are also available~\cite{Vincent,Zhang}. However, there is no model for the switching distribution in the intermediate current region for non-pulse waveforms. In the last case, the LLG equation has to be solved numerically.

\subsection{Previously Proposed STT-MTJ Based TRNGs}
\label{sec:related_work}

Emerging technologies, such as memristors, have been proposed for TRNGs that operate by applying a current pulse through the devices to randomly switch them with approximately 50\% probability~\cite{Jiang2017,TZhang,YandanWang}. The generated random number in these TRNGs is the state of the memristors at the end of operation.
Similarly, TRNGs based on STT-MTJ~\cite{Vatajelu,Fukushima,Oosawa,Qu2017,Wang2016} have been proposed.
In these designs, the current pulse is controlled by a feedback circuit in order to be robust to process variation and environmental changes. However, only Qu~\textit{et al.}~\cite{Qu2017} have analyzed the effects of process variation, and proposed to use several MTJ devices in parallel to mitigate those effects.

In \cite{Qu2018}, Qu \textit{et al.} proposed a differential approach, where two STT-MTJs are connected in series and in reverse orientation. A current pulse is driven through both of the MTJs simultaneously and a dedicated mechanism terminates the pulse when one of the MTJs is switched. The output bit is determined according to the end state of both MTJs. This design was shown to be robust to process variation, operating voltage, and temperature, due to the symmetry of the MTJs.

All of the aforementioned TRNGs use controlled current pulses to switch the MTJs.
To measure the duration of the current pulses, a clock with a period smaller than or equal to the pulse duration is needed. Additionally, the quality of the randomness will be influenced by the ability of the clock signal to accurately measure the pulse duration, further binding and complicating the system. Therefore, these TRNGs are incompatible with systems that have insufficient clock frequency or accuracy, such as low power systems with low frequency clock.

Lee \textit{et al.}~\cite{Lee2017} proposed to reduce the energy barrier between the P and AP of the MTJ to enable faster switching with reduced energy. This was achieved by using specially designed MTJ devices. Their proposed TRNG design uses multiple MTJs (to reduce the effect of process variation) with low energy barriers, which were set to an unstable state and released, letting the MTJs settle randomly to one of the stable states.
Vodenicarevic \textit{et al.}~\cite{Vodenicarevic} proposed a TRNG based on a similar approach, where the STT-MTJ's energy barrier between the P and AP states is sufficiently low to enable spontaneous switching in a reasonable time, without the aid of external stimuli and solely by thermal fluctuations. 
These two approaches require the use of specially designed MTJs due to the reduced energy barrier~\cite{Lee2017}, and, despite the advantages of reduced latency and energy, this approach makes the design more vulnerable to external magnetic fields, and thus to attacks.

Ghosh~\cite{Ghosh} comprehensively analyzed spintronics in security applications and showed that MTJ-based security circuits are sensitive to the effects of an external magnetic field.
Hence, it is essential to consider this effect when designing an MTJ based TRNG, to determine the security of such a design and the effects of nearby circuits.
However, none of the aforementioned STT-MTJ based TRNGs that rely on a current pulse~\cite{Vatajelu,Fukushima,Oosawa,Qu2017,Wang2016,Qu2018} were designed to account for the effects of an external magnetic field and the associated attacks.
The authors of~\cite{Lee2017,Vodenicarevic} only briefly consider the effects of an external magnetic field and their design relies on a special MTJ device that is more vulnerable to attacks. 
In this paper, we choose to use standard MTJ devices that are easier to fabricate in a standard process and have better robustness against attacks. We provide a comprehensive evaluation of the effects of an external magnetic field on the proposed TRNG design.

\section{Proposed TRNG Structure and Operation}
\label{sec:trng_struct}

\begin{figure}[tb]
\subfloat[]{\label{pic:TRNG_circuit_a} \includegraphics[width = 5.1cm,height=3.4cm]{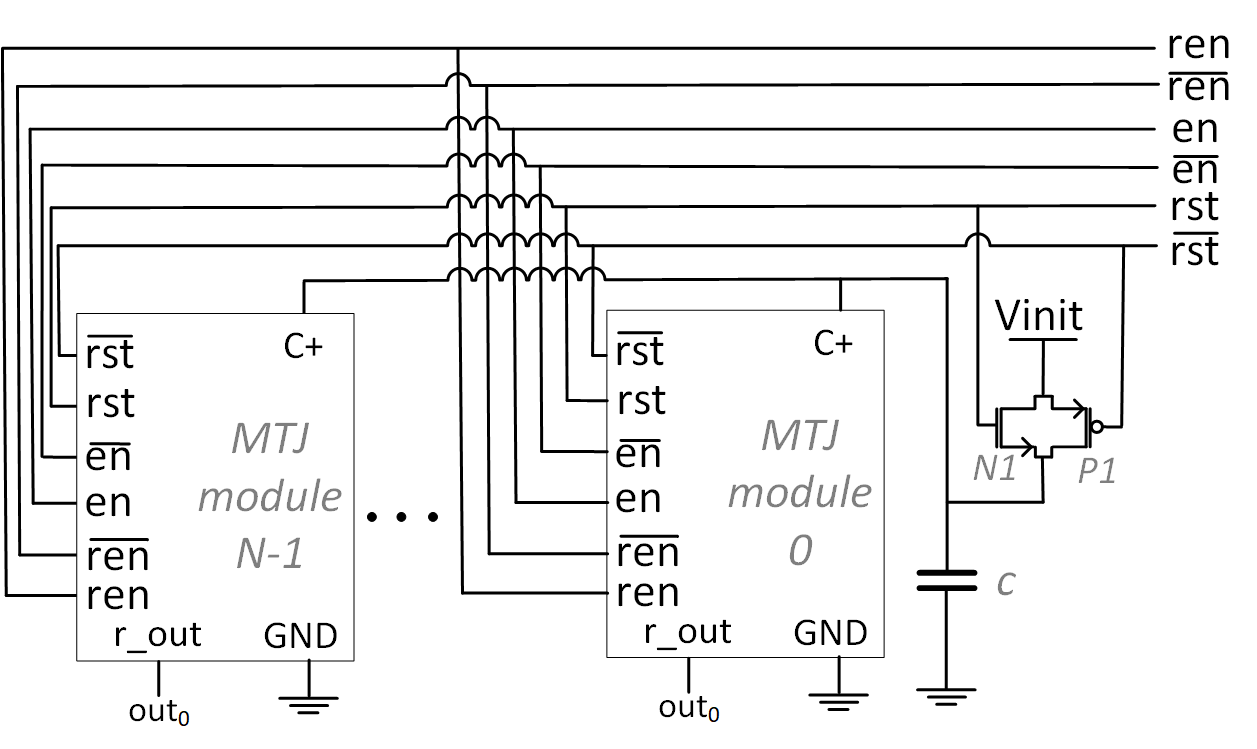}}
\subfloat[]{\label{pic:TRNG_circuit_b} \includegraphics[width = 3.5cm,height=3.4cm]{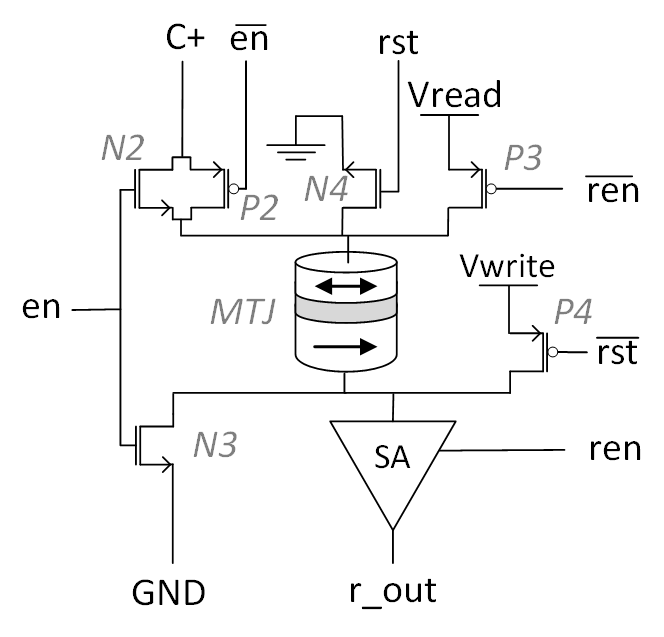}}
\caption{The proposed TRNG consists of (a) $N$ parallel connected MTJ-modules and a capacitor. (b) Schematic of the MTJ module.}
\label{pic:TRNG_circuit}
\vspace{-0.4cm}
\end{figure}

The proposed TRNG generates $N$-bit numbers and is composed of a capacitor, $N$ STT-MTJ devices, $N$ sense amplifiers, and transistors that serve as switches, as shown in Figure~\ref{pic:TRNG_circuit}. 

The TRNG operation consists of three steps, each taking a fixed amount of time. The first step, the \textit{Reset} step, charges the capacitor $C$ to the $V_{init}$ voltage (using transistors $N1$ and $P1$) and applies a current through the MTJ devices (using transistors $N4$ and $P4$), switching them all to the AP state.
The second step, the \textit{Enable} step, connects $C$ in parallel to all the MTJ devices (using transistors $N2, N3,$ and $P2$). This discharges $C$ through the MTJ devices, enabling them to switch to the P state with some probability. During the Enable step, the resistance of an MTJ drops if it is switched, making the capacitor discharge faster. This lowers but does not eliminate the switching probability of the other MTJs.
The third step, the \textit{Read} step, applies a small current through the MTJ devices (using transistor $P3$), and the sense amplifiers determine the state of each. The AP/P states are interpreted as '0'/'1' respectively. Overall, the TRNG outputs an $N$-bit word.  

The proposed TRNG relies on the stochastic switching time of the MTJ as its randomness source. Unlike previously proposed TRNGs, the randomness extraction operation in the Enable step is asynchronous and does not depend on a strict time measurement. The capacitor is sufficiently discharged during the Enable step to ensure a low probability for further switching until the end of the Read step. Hence, the randomness of the output does not change if the duration of the Enable step is longer than a certain lower bound. Thus, accurate measurement of the Enable step duration is not required. Note that although the Enable step is done asynchronously, the TRNG still uses a clock signal since a time measurement is still needed to transition between the operation steps.

\section{Evaluation}

\subsection{Measure for Randomness}
\label{subsec:measure_of_randomness}
Determining whether a sequence of numbers is random is considered difficult~\cite{NIST}. To try and overcome this problem, standard statistical test suites, such as the NIST SP 800-22~\cite{NIST}, are usually used to inspect for random properties. 
However, the proposed TRNG is evaluated using a simulation, which generates the TRNG outputs (N-bit words) in an i.i.d. (independent and identically distributed) manner. Each output is generated independently by numerically solving the stochastic differential equation system of the TRNG, which uses a different sequence of computer-generated random thermal noise for each TRNG output word. (For more information about and justifications for the simulation, see Section~\ref{subsec:sim}.) Hence, the simulation is designed such that there are no dependencies between bits from different TRNG outputs, but only the dependencies between bits in the same output word, reducing the dependency checks to within an output word.

To measure the dependencies between bits in the same output word, we use two measures: the Shannon entropy and the min-entropy of the output words. Entropy quantifies the amount of surprise in the outcome of the experiment. The higher the entropy, the more surprise in the experiment, \textit{i.e.} the experiment is more random. For an i.i.d. source with values from a finite set $\mathcal{X}$ with probability distribution function $p:\mathcal{X}\rightarrow[0,1]$, the Shannon entropy per word is $-\sum_{x\in\mathcal{X}}{p(x)\log_2p(x)}$ (with the definition that $0\cdot\log_20=0$) and the min-entropy per word is $\min_{x\in\mathcal{X}}(-\log_2p(x))$. Both entropies are measured in bits. If the number of elements in the finite set $\mathcal{X}$ is $m$, then both the Shannon and min entropies get a value in the range $[0,log_2m$], where $0$ entropy is achieved on the deterministic distribution and the $log_2m$ entropy is achieved on the uniform distribution. Taking $\mathcal{X}$ to be the set of N-bit words results in $log_2m=N$. Hence, the maximum entropy for an i.i.d. N-bit TRNG is $N$.

When there are dependencies between bits in the TRNG output, some output words will be more likely than others and the distribution of a single TRNG output word will deviate from the uniform distribution, resulting in a lower entropy than the maximum. Stronger dependency increases the deviation from the uniform distribution and the lowers the entropy. Hence, entropy is a measure for the dependencies in a single TRNG output word and an entropy close to maximum means low dependencies.

The min-entropy is a lower bound to the Shannon entropy (with equality achieved on the uniform and deterministic distributions) and it is the lowest amount of randomness a single sample of a random variable can give. Randomness extractors are sometimes designed to extract an output for every input, so the correct measure here is the min-entropy of their source~\cite{Salil}. The Shannon entropy is the expected randomness from a random variable. For an i.i.d. source, the Shannon entropy plays an important role in bounding the number of uniformly distributed bits that can be extracted from $n$ samples~\cite{Renner, Holenstein}.
Hence, the Shannon entropy gives us a notion of how many samples are required to extract a certain degree of randomness, while the min-entropy gives us the worst-case randomness of a single sample.  

\subsection{Simulation Methodology}
\label{subsec:sim}

\begin{table}[t]
\centering
\caption{Simulated Circuit Parameters. $R_{N2,N3,P2}$ is the modeled total effective resistance of transistors $N2,N3,P2$.}  
\vspace{-0.2cm}
\begin{tabular}{|p{3.5cm}|c||p{1cm}|c|} \hline
\rule{0pt}{2ex} Feature 							& Value 		&Feature 	& Value	\\ \hline
\rule{0pt}{2ex}NFET operation gate voltage			& $1.5 V$     	& $R_{on}$			& $1000 \Omega$	\\
PFET operation gate voltage							& $0 V$     	& $R_{off}$			& $2500 \Omega$	\\
$R_{N2,N3,P2}$, 2-bit TRNG						& $4450 \Omega$	& $V_{init}$		& $0.8 V$ \\
$R_{N2,N3,P2}$, 4-bit TRNG							& $3440 \Omega$	& $T_{enable}$		& $10 ns$\\
$R_{N2,N3,P2}$, 6-bit TRNG							& $2640 \Omega$	& $C$   			& $10 pF$\\
$R_{N2,N3,P2}$, 8-bit TRNG							& $1960 \Omega$	& Temp.				& $300K$\\
     		\hline
\end{tabular}
\label{table:circuit_details}
\vspace{-0.4cm}
\end{table}
We evaluated our TRNG with Monte-Carlo simulations for the Enable step for different topologies, each with a different number of MTJ devices (different \textit{N}).
The simulation numerically solves the differential equation system of the MTJs (stochastic LLG equations) and the capacitor. The LLG equations are used since a non-pulse current waveform is passed through the MTJ devices. 
The transistors $N2,N3,$ and $P2$ were modeled by a constant resistance. The equations were solved using a standard midpoint scheme~\cite{d’Aquino} assuming no external magnetic field (unless otherwise stated) and the stochastic term was interpreted in the sense of Stratonovich. We could not obtain an analytic expression for the TRNG output distribution.

Each iteration of the Monte-Carlo simulation produces the TRNG output binary word. For each measurement of entropy, 2000 iterations were conducted. The probability of each TRNG output was evaluated as its frequency of appearance. However, when the parameters of the simulated MTJs were identical (\textit{i.e.}, with no device-to-device variations), the probability was evaluated as the frequency of the corresponding Hamming weight divided by the number of outputs with the same Hamming weight\footnote{From symmetry, outputs with the same Hamming weight have the same probability.}, thus increasing the accuracy.
Note that each iteration of the Monte-Carlo simulation produces a single $N$-bit output of the TRNG, meaning that 
the simulation produces the probability of a single output of the TRNG. 

When this simulation model is used to extract the entropy of the TRNG, the TRNG is assumed to be an i.i.d. source, since the model does not include the correlation between consecutive runs of the TRNG. This assumption is justified since the MTJs are always in the AP state prior to the Enable step, regardless of the output of the last run. Furthermore, thermal fluctuations occur constantly; hence, the exact position of the magnetization (around the AP state) at the beginning of the Enable step is itself random. This results in a fresh start in every run.

Nevertheless, consecutive runs will be correlated in a real-world TRNG. Some correlation will be caused by changes in the operation parameters but not by true causality between samples. If, for example, during the sampling of the TRNG, a nearby circuit will periodically work with cycle time $T$, the magnetic field on the MTJs will change with the same period. As a result, samples with time $T$ between them will show correlation just because they have the same distribution, not because they have true causality. Similar effects can come in the form of temperature, voltage, \textit{etc.}

The STT-MTJ devices are modeled as device C from~\cite{Vincent}, a standard in-plane STT-MTJ that can be used for memory design and has the lowest switching current in~\cite{Vincent}. The circuit parameters are listed in Table~\ref{table:circuit_details}. Different values for the modeled effective resistance of transistors $N2, N3,$ and $P2$ were simulated for each topology, and were chosen to maximize the entropy; results are shown in Table~\ref{table:circuit_details}. To find the size of the transistors and verify the accuracy of the constant resistance model, we performed circuit simulations (with resistors instead of MTJs) in Cadence Virtuoso using a 28nm GlobalFoundries process. The Virtuoso simulations showed that the transistors transition quickly at the beginning of the Enable step. Since this transition time is faster than the switching time of the MTJs, it can be ignored. During the rest of the Enable step the total resistance of the transistors is approximately constant, verifying that our constant resistance model of the transistors is acceptable for the evaluation of our design.

Since the resistance of the interconnect is a few ohms per $\mu m$ while the resistances of the transistors and MTJs are in the order of $k\Omega$, we neglect the wire resistance. The difference in resistance due to difference in wire lengths between the MTJ modules is also in the order of a few ohms and is well within the process variation considerations in Section~\ref{subsubsec:proc_var}. Hence, it is neglected as well. Additionally, we measured the parasitic capacitance and leakage currents in our design. The simulation shows that the parasitic values are several orders of magnitude lower than the non-parasitic values, and hence they are ignored. 

\begin{figure}[tb]
\includegraphics[width = 8.5cm,height=8.3cm]{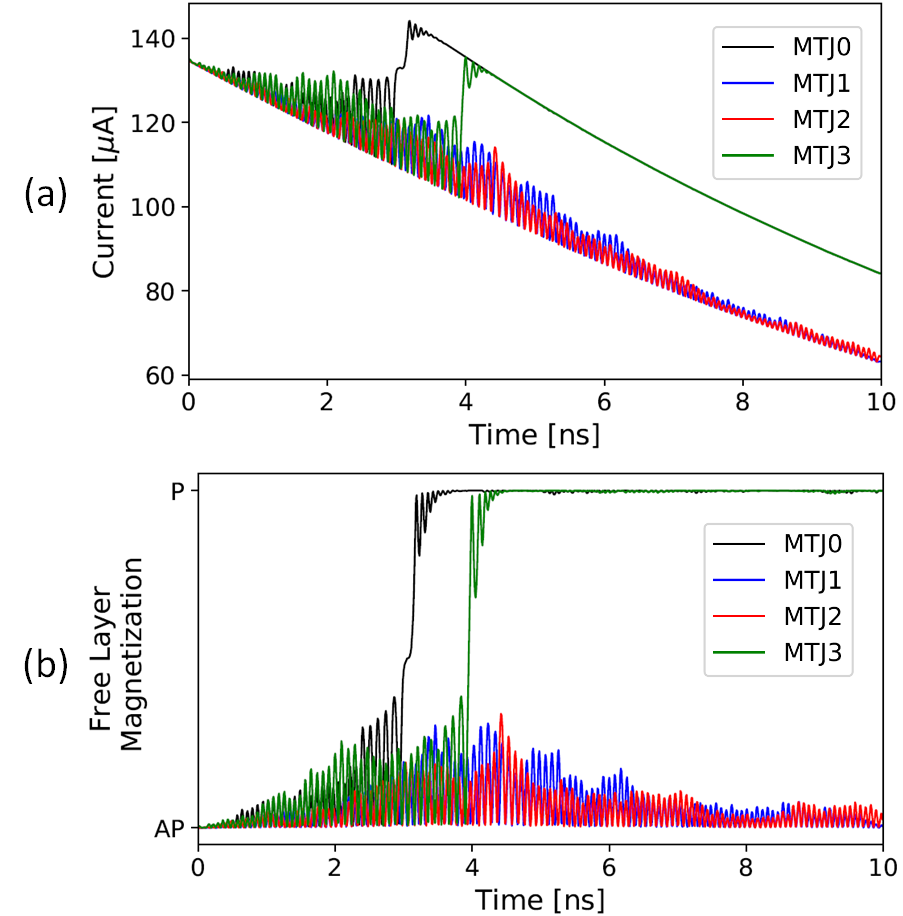}
\vspace{-0.1cm}
\caption{An example for a single iteration of the simulation for $N=4$. (a) The current through the MTJs during the Enable step. (b) The magnetization direction of the free layer in the dimension of the fixed layer magnetization during the Enable step. In this example MTJ0 and MTJ3 are flipped to the P state, and the TRNG output word will be $1001$.}
\label{fig:iteration}
\vspace{-0.28cm}
\end{figure}

The number of MTJ devices in the design, \textit{i.e.} $N$, was restricted to 2, 4, 6, and 8 because the larger $N$ is, the shorter the discharge time of the capacitor. Hence, a stronger current should flow through the MTJ devices to maintain the same switching probability, requiring a lower transistor resistance. The lower resistance will shorten the discharge time further, but the resulting switching probability of the MTJs will increase. However, a lower transistor resistance means a larger transistor size. A larger $N$ can improve the performance of the TRNG (see Section~\ref{subsec:eval_performance}). However, the simulation was restricted to $N\le8$ since the transistor sizes required for the $N=8$ topology are considered large (width of a few hundred nano-meters ). A larger $N$ can be achieved by considering different system parameters (\textit{e.g.}, higher $V_{init}$, larger capacitor).

An example for a single iteration of the simulation with $N=4$ is shown in Figure~\ref{fig:iteration} ($N=4$ was chosen for clarity). The currents through the MTJs during the simulated Enable step are shown in parallel to the magnetization of the free layer (in a single dimension, the dimension of the fixed layer magnetization). When an MTJ switches to the P state, its resistance drops and the current through it increases. When the current reaches the P state, the current through the MTJ and the magnetization of the free layer remain steady, since the current through the MTJ continues to direct the state to the P state.

We compare our results to CMOS-based TRNGs. Since previously proposed STT-MTJ based TRNGs require a high frequency clock or a modified STT-MTJ device and do not include process variation or external magnetic field influence in their evaluation, we have not compared our results to those designs. 

\subsection{NIST Statistical Test Suite}

The National Institute of Standards and Technology (NIST) SP 800-22 rev.1a~\cite{NIST} test suite is commonly used to evaluate random number generators. The suite is composed of several statistical tests, each operating on a string of bits. The test indicates whether this string of bits is likely to come from a uniform i.i.d. source. To conduct the test suite, a number of bit sequences are retrieved from the generator to be tested (sequences do not share bits between them). 
Each test is run on all sequences, resulting in a P-value for each sequence on each test. The P-value quantifies the distance between the test result for that sequence and the expected result for a uniform i.i.d. sequence. A sequence is said to pass a test if the P-value for that sequence and test is above a threshold. Then two scores are given for each test, a success rate and a P-value. The success rate is the proportion of sequences that pass that test, while the P-value is a number between 0 and 1. This number quantifies the distance of the sequences' P-value distribution from the expected P-value distribution of a uniform i.i.d. source for that test. Note that even a perfect source will not produce a perfect success rate and P-value (\textit{i.e.} 1) for a test, due to statistical deviations.
A generator is said to pass a test if the test's success rate and P-value are both above a threshold.

The proposed TRNG does not have a uniform distribution since output words with different Hamming weight have different probabilities, while output words with the same Hamming weight have the same probability (due to symmetry). Since the design was calibrated for maximum entropy, the words with minimum (all zeros) and maximum (all ones) Hamming weights have the lowest probabilities, while the words with the half zeros and half ones have the highest probability. This results in a bias towards certain bit patterns when a sequence of output words is considered as a sequence of bits. Hence, the NIST test suite is \textit{expected to fail} on the proposed TRNG without post-processing. Failing the NIST test suite does not mean that the TRNG is not random; it means only that the TRNG distribution does not appear to be uniformly distributed. Furthermore, non-uniformity does not mean that the TRNG cannot be used for cryptography. It means only that post-processing might be a prerequisite for this use. (Post-processing might be used regardless, to compensate for process variation and interference effects~\cite{Kwok2011,BHUNIA2019}.)

To show that the proposed TRNG is sufficiently random, the TRNG output words are post-processed by a simple and reversible function before evaluation by the test suite. The reversibility of the processing indicates that the new bit sequence has the same information as the original sequence. To define the post-processing, the TRNG output word sequence is denoted by $\{w_1,w_2,w_3,\ldots\}$ and the processed word sequence is denoted by $\{z_1,z_2,z_3,\ldots\}$, each of which are $N$-bit words. Then, the post-processing is defined as $z_1=w_1$ and $z_i=w_i\oplus z_{i-1}$ for $i>1$ ($\oplus$ is the bit-wise XOR operation). The bit sequence is taken as the bits of $\{z_i\}_{i\ge1}$, starting from $z_1$, in a big-endian fashion. This operation is reversible (since $w_1=z_1$ and $w_i=z_i\oplus z_{i-1}$ for $i>1$) and simple to implement (requiring an $N$-bit register and $N$ XOR gates). Note that this post-processing is not a qualified cryptographic hash function or randomness extractor. It is used solely to demonstrate that the TRNG has sufficient randomness to pass the test in the NIST test suite.

For each topology of the TRNG, $1024$ sequences of $1024$ bits each (a total of $2^{20}$ bits) were generated for the NIST test suite. The results of the P-value and success rates for the tests with their thresholds are listed in Table~\ref{table:NIST}. To perform all of the tests in the suite, a substantially larger number of bits is needed, which will require an unfeasible amount of time to produce with the TRNG simulation. Therefore, the tests listed in Table~\ref{table:NIST} are those that can be run with the produced number of bits.
All topologies passed all executed tests.

\begin{table}[t]
\centering
\caption{Results of the NIST test suite~\cite{NIST} for the proposed TRNG for different numbers of bits (N). P-value threshold is .0001 and success rate threshold is $1004/1024\approx$ .980 for testing 1024 sequences. The proposed TRNG passed all the tests.}  
\vspace{-0.2cm}
\begin{tabular}{|@{\,\,}c@{\,\,}|@{\,\,}c@{\,\,}|@{\,\,}c@{\,\,}|@{\,\,}c@{\,\,}|@{\,\,}c@{\,\,}|@{\,\,}c@{\,\,}|@{\,\,}c@{\,\,}|@{\,\,}c@{\,\,}|@{\,\,}c@{\,\,}|} \hline
\setlength{\tabcolsep}{2pt}
\rule{0pt}{2ex}  & \multicolumn{2}{@{\,\,}c|@{\,\,}}{$N = 2$} & \multicolumn{2}{@{\,\,}c|@{\,\,}}{$N = 4$} & \multicolumn{2}{@{\,\,}c|@{\,\,}}{$N = 6$} & \multicolumn{2}{@{\,\,}c@{\,\,}|}{$N = 8$}	\\ 
\makecell{Test\\Name}    & \makecell{P\\value}   & \makecell{Succ.\\Rate}         & \makecell{P\\value} 	& \makecell{Succ.\\Rate}         & \makecell{P\\value} 	& \makecell{Succ.\\Rate}         & \makecell{P\\value} 	& \makecell{Succ.\\Rate}\\  \hline\hline
\rule{0pt}{2ex}Frequency	                    & .28  & .993 & .0079    & .992 & .69      & .992 & .54      & .985\\ \hline
\makecell{Block\\Frequency}     & .73  & .995 & .43      & .990 & .53      & .985 & .20      & .988\\ \hline
\rule{0pt}{2ex}Runs 	                        & .038 & .990 & .22      & .988 & .068     & .983 & .25      & .992\\ \hline
\rule{0pt}{2ex}\makecell{Longest\\Run}         & .87  & .990 & .34      & .993 & .073     & .990 & .28      & .990\\ \hline
\rule{0pt}{2ex}Serial    (1)                   & .17  & .988 & .029     & .982 & .081     & .989 & .14      & .985\\ \hline
\rule{0pt}{2ex}Serial    (2)                   & .52  & .994 & .17      & .987 & .44      & .989 & .58      & .987\\ \hline
\makecell{Approximate\\Entropy\footnote{This is not the same entropy as computed in the rest of the paper; see~\cite{NIST} for details.}}                     & .28  & .982 & .00078   & .984 & .75      & .984 & .065     & .986\\ \hline
\rule{0pt}{2ex}Cusum     (1)                   & .026 & .990 & .0011    & .990 & .016     & .988 & .00067   & .983\\ \hline
\rule{0pt}{2ex}Cusum     (2)                   & .16  & .992 & .0082    & .992 & .0014    & .989 & .055     & .985\\ \hline

\end{tabular}
\vspace{-0.4cm}
\label{table:NIST}
\end{table}

\subsection{Entropy per Output}
\label{subsec:eval_performance}
We evaluated the  entropy of the TRNG for different design, environmental, and process parameters. 

\begin{figure}[tb]
\includegraphics[width = 8.5cm,height=8cm]{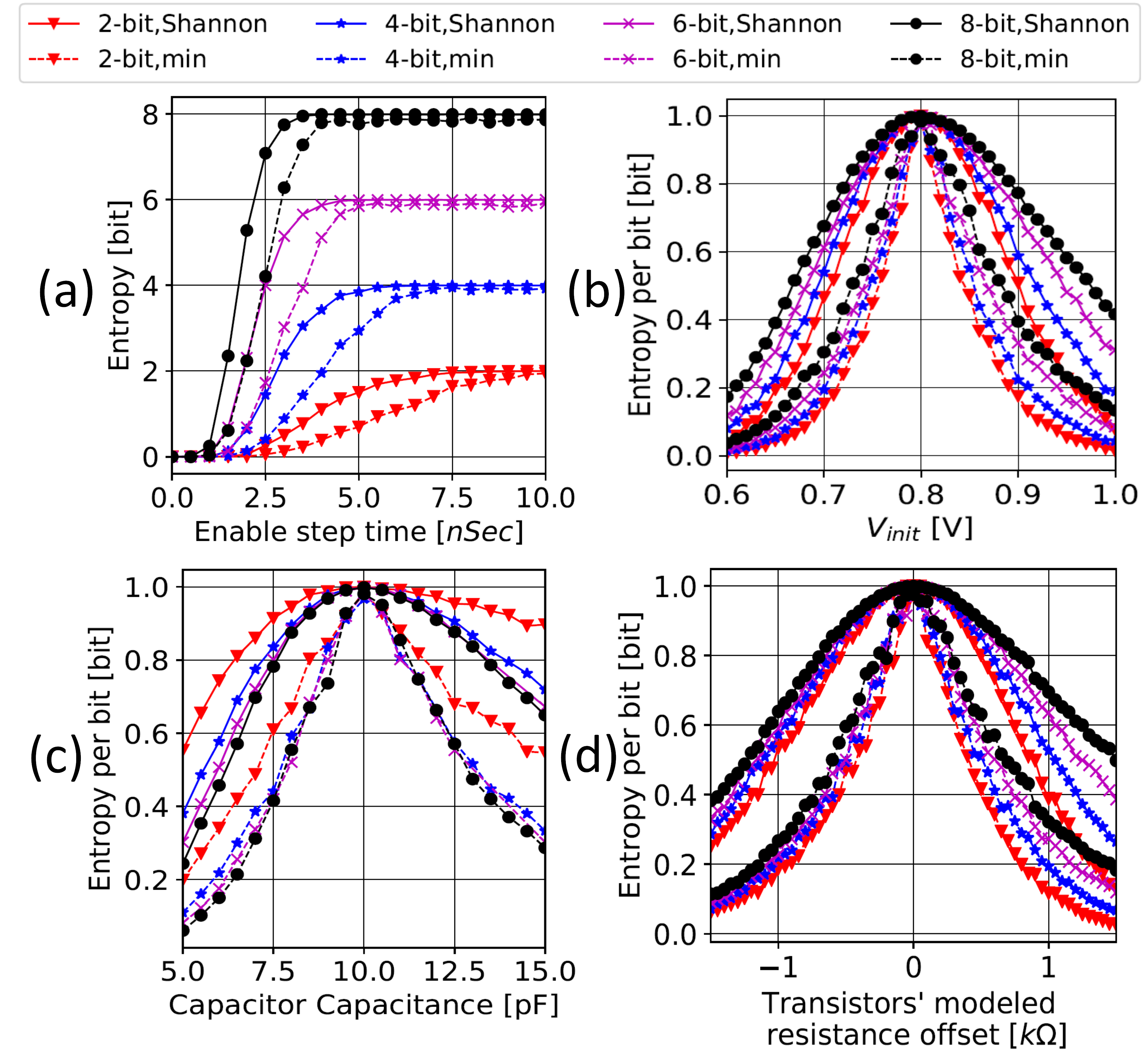}
\vspace{-0.1cm}
\caption{Entropy of the TRNG for different (a) enable step time duration, (b) $V_{init}$, (c) $C$, and (d) offset in the total effective resistance of transistors $N2,N3,P2$.}
\label{fig:design_param}
\vspace{-0.28cm}
\end{figure}

\subsubsection{Design Parameters}
\label{subsubsec:design_param}

Include the Enable step duration time, $V_{init}$, $C$, and the effective modeled resistance of $N2, N3,$ and $P2$.
Simulation results of the entropy for different design parameters are shown in Figure~\ref{fig:design_param}.
The design parameters trade-off between the different performance measures of the TRNG (entropy, operation time, area and power) while using the same MTJ devices.
An important observation is that our design can, in the ideal case, reach nearly the maximum possible entropy (1-bit entropy per MTJ device).

Figure~\ref{fig:design_param}(a) shows the effect of the Enable step duration, as mentioned in Section~\ref{sec:trng_struct}. When the duration is sufficiently long, the randomness of the TRNG is maximal. This allows the TRNG to be used in low-frequency devices, where the time measurement has a low resolution.  

It is evident that a small change in $V_{init}$, from its designed value of $0.8V$, can change the entropy. The value of the initial capacitor voltage affects the duration of the Reset step and the entropy throughput of the TRNG (see Section~\ref{subsec:trhoughput}). However, reasonable variations in the capacitance of the capacitor (less than $0.5pF$, or approximately 5\%) have little effect on the entropy, since the capacitance is relatively large.

Additionally, we can conclude that modeling the open transistors by a constant resistance is tolerable.
Deviations in the range of $250\Omega$ from the designed value do not reduce the entropy much. From the Cadence Virtuoso simulation (see Section~\ref{subsec:sim}) we verified that the actual effective resistance of $N2, N3,$ and $P2$ fluctuates well within the range of $\pm 250\Omega$ around the designed value. 

\subsubsection{Environmental Parameters}
\label{subsec:eval}

\begin{figure}[tb]
\centering
\includegraphics[width = 8.8cm,height=10.0cm]{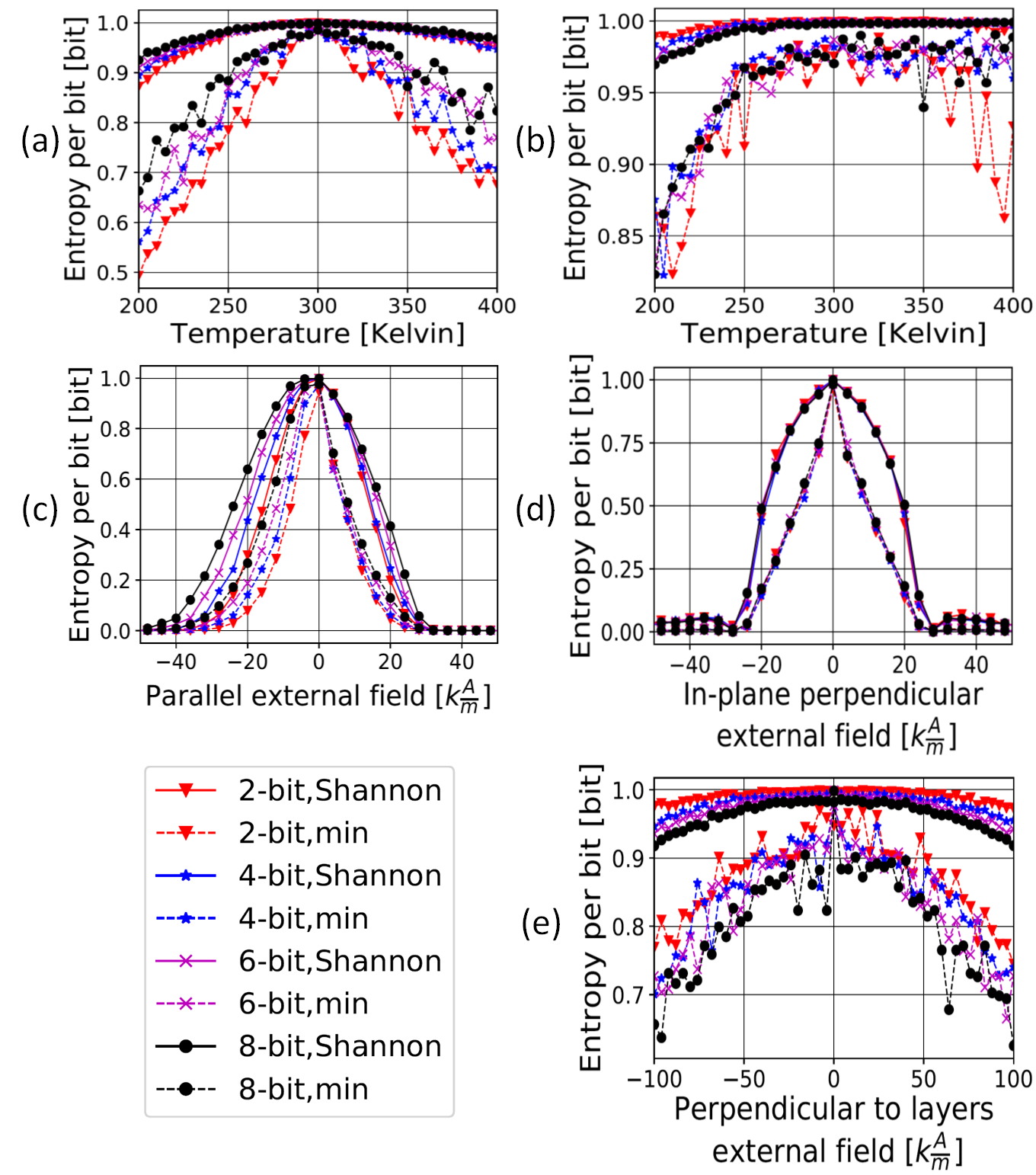}
\caption{(a)-(b) Effect of temperature: (a) a non-changing MTJ resistance, (b) including a change in MTJ resistance, where $R_{on}$ is constant and $TMR=\frac{R_{off}-R_{on}}{R_{on}}$ changes at a linear rate of $-0.4\frac{\%}{^{\circ}K}$. (c)-(e) Effect of a constant external magnetic field (fixed direction and varying magnitude): (c) parallel to the fixed layer magnetization, (d) in-plane and perpendicular to the fixed layer magnetization, (e) perpendicular to the layer's plane.}
\label{fig:extrnal_param}
\vspace{-0.5cm}
\end{figure}

These parameters are external to the TRNG and can be altered by an adversary. (see Section~\ref{sec:adversary}).
The effect of temperature and the external magnetic field on the MTJ devices are considered here, and their influence on the entropy is shown in Figure~\ref{fig:extrnal_param}. 

In the LLG equation, the temperature affects only the thermal fluctuations of the MTJ. However, the resistance of the MTJ is temperature dependent ~\cite{Drewello2008,WANG2015}. While the resistance of the MTJ in the P state is roughly constant, the resistance of the AP state changes more considerably with the temperature~\cite{Drewello2008}. The $TMR=\frac{R_{off}-R_{on}}{R_{on}}$  changes in an approximately linear manner around $300^{\circ}K$, with a rate of $-0.2\frac{\%}{^{\circ} K}$ to $-0.4\frac{\%}{^{\circ} K}$ according to~\cite{Drewello2008}. To investigate the temperature dependent behavior of the proposed TRNG design, we simulated different temperatures while maintaining $R_{on}$ constant, and changing  $R_{off}$ to produce a $TMR$ change with a linear rate $0\frac{\%}{^{\circ} K}$ (\textit{i.e.,} constant $TMR$) (Figure~\ref{fig:extrnal_param}(a)) and $-0.4\frac{\%}{^{\circ} K}$ (Figure~\ref{fig:extrnal_param}(b)).

If the $TMR$ rate is $0\frac{\%}{^{\circ} K}$, only the thermal fluctuations are affected by the temperature change. In this case, a degradation in the TRNG entropy is evident when the temperature deviates from $300^{\circ}K$. 
The switching probability of the MTJs changes with the temperature (as can be seen by the min-entropy in Figure~\ref{fig:extrnal_param}(a) and as reported in other works~\cite{WANG2015,Vatajelu2016,Khan2018}), but the Shannon-entropy of the TRNG output word remains high in the examined temperature range.

If the $TMR$ rate is $-0.4\frac{\%}{^{\circ} K}$, the entropy of the TRNG is better than in the $0\frac{\%}{^{\circ} K}$ case. When the temperature is lower (higher) than $300^{\circ}K$, the resistance of $R_{off}$ increases (decreases), resulting in a lower (higher) initial current through the MTJs but with a longer (shorter) discharge time for the capacitor. The longer (shorter) discharge time results in a longer (shorter) time for a non-negligible switching probability current, which might negate the reduced (increased) switching probability produced by the thermal fluctuations. These two effects, the thermal fluctuations and the current waveform through the MTJ, interact in a non-trivial way with the entropy due to their different non-linear characteristics.

This analysis shows that the Shannon entropy of the proposed TRNG design behaves well under temperature changes.

\begin{figure}[tb]
\subfloat[min-entropy]{\label{fig:rotate_min} \includegraphics[width = 3.4cm,height= 3.32cm]{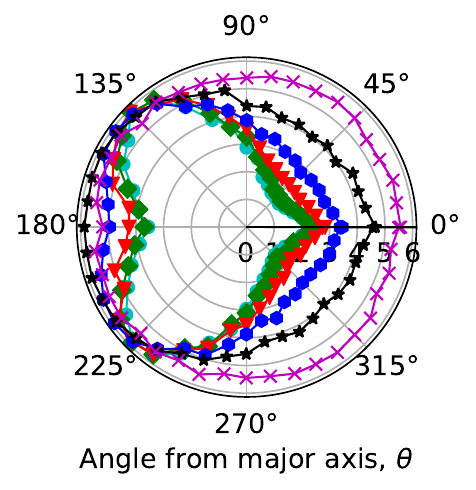}}
\captionsetup[subfigure]{oneside,margin={1.9cm,0cm}}
\subfloat[Shannon entropy]{\label{fig:rotate_shannon} \includegraphics[width = 5.1cm,height= 3.36cm]{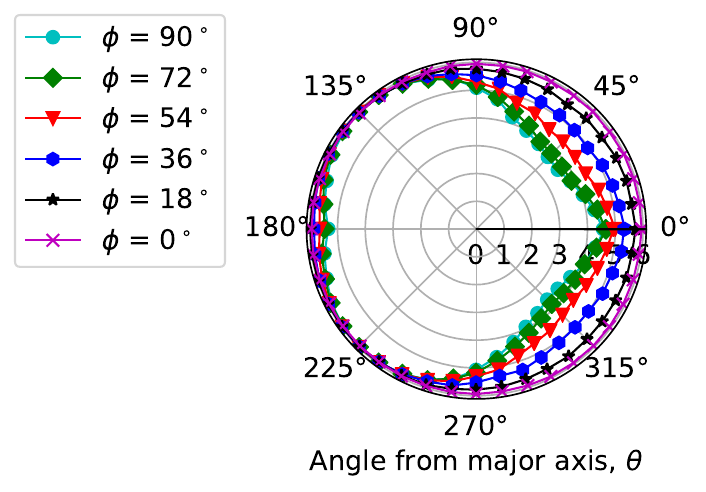}}
\caption{Effect of a constant external field in different angles and fixed magnitude of 10 $k\frac{A}{m}$ on the entropy of a 6-bit TRNG. $\phi$ is the angle of the field from the axis perpendicular to the MTJ plane, $\theta$ is the in-plane angle from the magnetization of the fixed layer.}
\vspace{-0.6cm}
\label{fig:rotate}
\end{figure}

When considering the external magnetic field, a sufficiently high field can reduce the entropy to $0$, as shown in Figure~\ref{fig:extrnal_param}(b)-(c). Every device using MTJs will be susceptible to a strong enough magnetic field, which can be exploited by an adversary (see Section~\ref{sec:adversary} for further discussion). When the external magnetic field is perpendicular to the fixed layer magnetization (the fixed layer magnetization of all MTJs are parallel), from the symmetry of the MTJ the entropy is expected to be symmetrical around zero magnetic field, as confirmed in Figures~\ref{fig:extrnal_param}(c) and~\ref{fig:extrnal_param}(d).
Additionally, since an in-plane MTJ is used, the external field perpendicular to the MTJ layers has little influence on the entropy. Therefore, when designing the TRNG circuit, we would like to position nearby wires in the same plane as the TRNG, and use vias as short as possible and positioned as far from the TRNG as possible.

\begin{figure}[tb]
\captionsetup[subfigure]{oneside,margin={0cm,0cm}}
\subfloat[min-entropy]{\label{fig:rotateAC_min} \includegraphics[width = 3.4cm,height= 3.32cm]{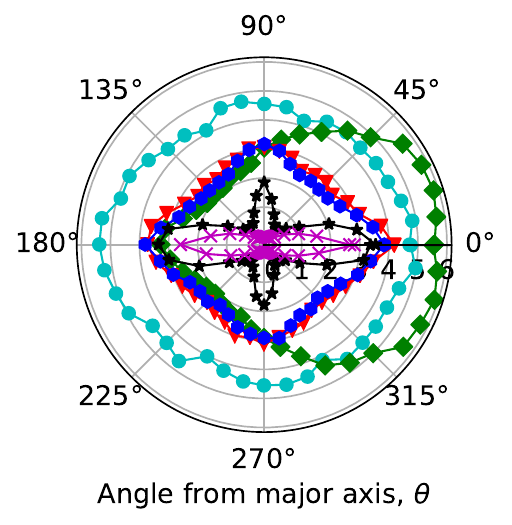}}
\captionsetup[subfigure]{oneside,margin={1.9cm,0cm}}
\subfloat[Shannon entropy]{\label{fig:rotateAC_shannon} \includegraphics[width = 4.9cm,height= 3.36cm]{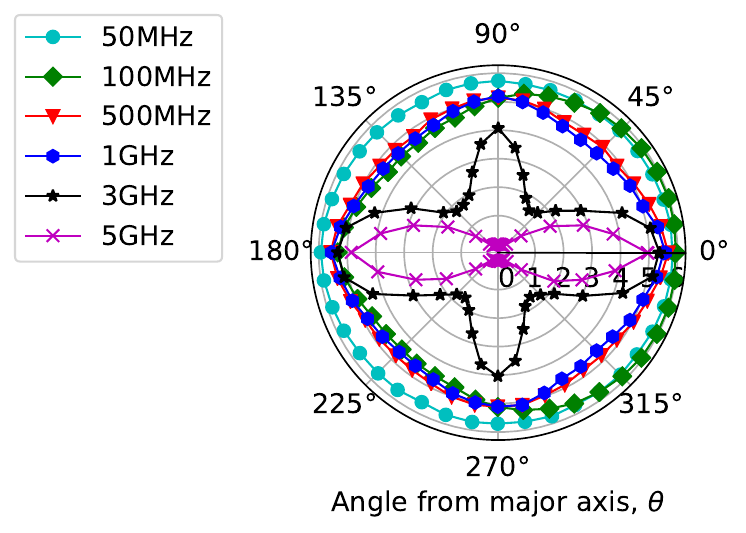}}
\caption{Effect of an alternating external field on multiple in-plane directions and fixed magnitude of 10 $k\frac{A}{m}$ on a 6-bit TRNG.}
\label{fig:rotateAC}
\end{figure}

To further consider an external magnetic field, we simulated the effect for other directions of the field relative to the fixed layer magnetization of the MTJs.
Figure~\ref{fig:rotate} shows the effect of the direction of an external field with a magnitude of $10k\frac{A}{m}$ on a 6-bit TRNG. The worst effect is for approximately $\theta= 45^\circ\backslash315^\circ$, rather than in the direction of the fixed-layer magnetization ($\theta= 0^\circ\backslash 180^\circ$) or perpendicular to it ($\theta=90^\circ\backslash 270^\circ$).
Figure~\ref{fig:rotateAC} shows how an alternating external field applied from multiple directions, with a fixed magnitude of $10k\frac{A}{m}$ and different frequencies, affects a 6-bit TRNG. Since a field perpendicular to the MTJ affects the entropy much less than other directions, the alternating field simulation was limited only to the in-plane directions. Note that the Enable step duration is $10ns$ (Table~\ref{table:circuit_details}), which corresponds to a $100MHz$ frequency. 

An interesting result is that the performance under an alternating external field in the range of $50MHz$ to $1GHz$ is actually better than the performance under a constant external field. This implies that nearby circuits at frequencies up to $1GHz$ will induce minimal performance loss. Another observation is that an alternating external field above a certain frequency is very effective. Countermeasures against external fields are discussed in Section~\ref{sec:adversary}.

\subsubsection{Process Variation}
\label{subsubsec:proc_var}

Process variation in the TRNG will result in different switching probabilities for the different MTJs, biasing some to switch with higher probabilities and some with lower probabilities, resulting in bias for certain output words. However, the i.i.d. property of the TRNG is not affected by process variation (see Sections~\ref{subsec:sim}): it affects only the distribution of an output word but not the dependencies between different output words. Hence, the methodology presented in Section~\ref{subsec:measure_of_randomness} is used in this section as well.

Variations in the MTJs and in the transistors operating in the Enable step have been considered. Section~\ref{subsubsec:design_param} shows that reasonable variation in capacitor $C$ does not change the entropy; hence, $C$ is not considered for process variation. For the transistors, we modeled variations in their fixed resistance. For the MTJ, we modeled variations in the physical size of the devices: major and minor axis length (the MTJ shape is an ellipse cylinder~\cite{Vincent}), the thickness of the free layer, and the thickness of the oxide layer (tunnel barrier layer). We generated 1000 different instances for each TRNG topology and their entropy was evaluated. The parameters were drawn independently from a Gaussian distribution with mean as the designed value (listed in Table~\ref{table:circuit_details} and in~\cite{Vincent}) and a standard deviation of $5\%$~\cite{JLi}.

The geometry of the MTJ affects its demagnetization factors~\cite{Vincent} and resistance. The method presented in~\cite{Goode} was used to compute the new demagnetization factors. The MTJ resistance is proportional to the exponent of the oxide layer thickness ($t_{ox}$) and inversely proportional to its area ($A$)~\cite{JLi}, \textit{i.e.}, $R_{on},R_{off}\propto \frac{e^{\rho\cdot t_{ox}}}{A}$ ($\rho$ is a constant that depends on the device technology). The thickness of the oxide layer appears in the simulation only as part of the MTJ resistance. Since the oxide layer thickness and the coefficient $\rho$ are unavailable for the simulated device, the process variation of the oxide layer was evaluated as an additional variation in the MTJ resistance by a Gaussian distribution with a standard deviation of $5\%$. The simulation results are listed in Table~\ref{table:proc_var}. The results show that the entropy per bit increases with the number of MTJs, resulting in a twofold increase in the TRNG total entropy.

Even though most TRNG instances will have sufficiently high entropy under process variation, some TRNG instances might still produce low entropy and be unusable. To protect the TRNG from such an event, several TRNG instances should be fabricated together. Fortunately, the largest area is consumed by the capacitor $C$ and the sense amplifier (see Section~\ref{subsec:area}), which can be shared among the different fabricated MTJ modules (the implications of this suggestion are not discussed in this paper). Thus, the process variation robustness can be increased with a low area overhead. 

\subsection{Entropy Generating Rate}
\label{subsec:trhoughput}

In many systems, a large number of entropy bits are required. In this case, the entropy generating rate (entropy bits per second) is the desired performance metric. Nevertheless, the TRNG is not required for the entire operation of the system, since there is no need for random words most of the time.
The generating rate gives a notion of the delay time between starting the TRNG and having the desired number of entropy bits, enabling the system to react faster. Since many TRNG generated numbers are involved, the generating rate refers to the Shannon entropy.
The entropy generating rate is measured in units of \textit{entropy bits per second}, which is the amount of entropy produced by the TRNG in a second.

\begin{table}[t]
\centering
\caption{Entropy results with process variation showing the average, standard deviation, median, and the 10th percentile}  
\vspace{-0.2cm}
\begin{tabular}{|c|c|c|c|c|c|c|c|c|} \hline
\rule{0pt}{2ex} 	&   \multicolumn{4}{|c|}{Shannon Entropy per Bit}  		& \multicolumn{4}{|c|}{Min-Entropy per Bit}	\\
\rule{0pt}{2ex}	$N$		& Avg. 	& sd 	& Med. 	& $P_{10}$			& Avg. 	& sd 	& Med. 	& $P_{10}$\\ \hline\hline
\rule{0pt}{2ex} 2				& 0.74 	& 0.19	& 0.76	& 0.46				& 0.46	& 0.21	& 0.47	& 0.17	\\ \hline
\rule{0pt}{2ex} 4 				& 0.79	& 0.12	& 0.80	& 0.64				& 0.51	& 0.14	& 0.51	& 0.33  \\ \hline
\rule{0pt}{2ex} 6				& 0.82 	& 0.08	& 0.83	& 0.72				& 0.54	& 0.10	& 0.55	& 0.41	\\ \hline
\rule{0pt}{2ex} 8\footnote{For the 8-bit TRNG we simulated 6000 iterations.}				& 0.86 	& 0.06	& 0.86	& 0.78				& 0.58	& 0.08	& 0.59	& 0.47	\\ \hline
\end{tabular}
\vspace{-0.4cm}
\label{table:proc_var}
\end{table}

To determine the entropy generation rate, all three operation steps should be considered.
The Reset step duration is dominated by the capacitor charging time.
If we model the passgate $P1-N1$ connecting the capacitor to $V_{init}$ (Figure~\ref{pic:TRNG_circuit}) as a resistor of $1.5K\Omega$, the capacitor charging time from $0V$ to $0.79V$ (98.8\%) is $66ns$.
The duration of the Enable step is $10ns$ (Table~\ref{table:circuit_details}).
In the Read step, the states of the MTJs are read using sense amplifiers. If we take the read latency of $2.8ns$ reported by~\cite{Dong}, an output is produced every $78.8ns$.

For an 8-bit TRNG with process variation, $90\%$ of instances have an entropy generation rate between $79.2Mbps$ and $101.5Mbps$. This rate can be improved by terminating the Enable step earlier (as seen in Figure~\ref{fig:extrnal_param}(a)) and by reducing the charging time (since the capacitor is not fully discharged immediately after an operation). For an 8-bit TRNG, the entropy generation rate can be improved to $99.7-127.8Mbps$ for $90\%$ of the instances. However, the actual generation rate of our proposed design depends on the system clock used to measure the duration of the steps; the computed times represent the best case.

\subsection{Area and Energy}
\label{subsec:area}
To estimate the area of the TRNG, we evaluate the area for each component in the proposed circuit.
The area of a single STT-MTJ device is $0.003\mu m^2$~\cite{Vincent}. To find the capacitor $C$ area, we modeled it as a MOS capacitor with GlobalFoundries 28nm technology and obtained an area of $400\mu m^2$ using Cadence Virtuoso.
We evaluated the sense amplifier area from~\cite{Dong} as the summed area of its transistors.
All transistors were evaluated with a width of $500nm$ and minimum length (transistor sizes are not specified in~\cite{Dong}).
(This size upper bounds the area of the transistors shown in Figure~\ref{pic:TRNG_circuit}.)
The resulting area of an MTJ module is $0.395\mu m^2$.
The sense amplifier in~\cite{Dong} uses an additional capacitor, but its area is relatively small since the read duration is $2.8ns$. Hence, each MTJ module area was approximated as $0.6\mu m^2$. Table~\ref{table:cmpr} lists the TRNG area for different $N$.

\begin{figure}[tb]
\centering
\includegraphics[width = 8.5cm,height=5.7cm]{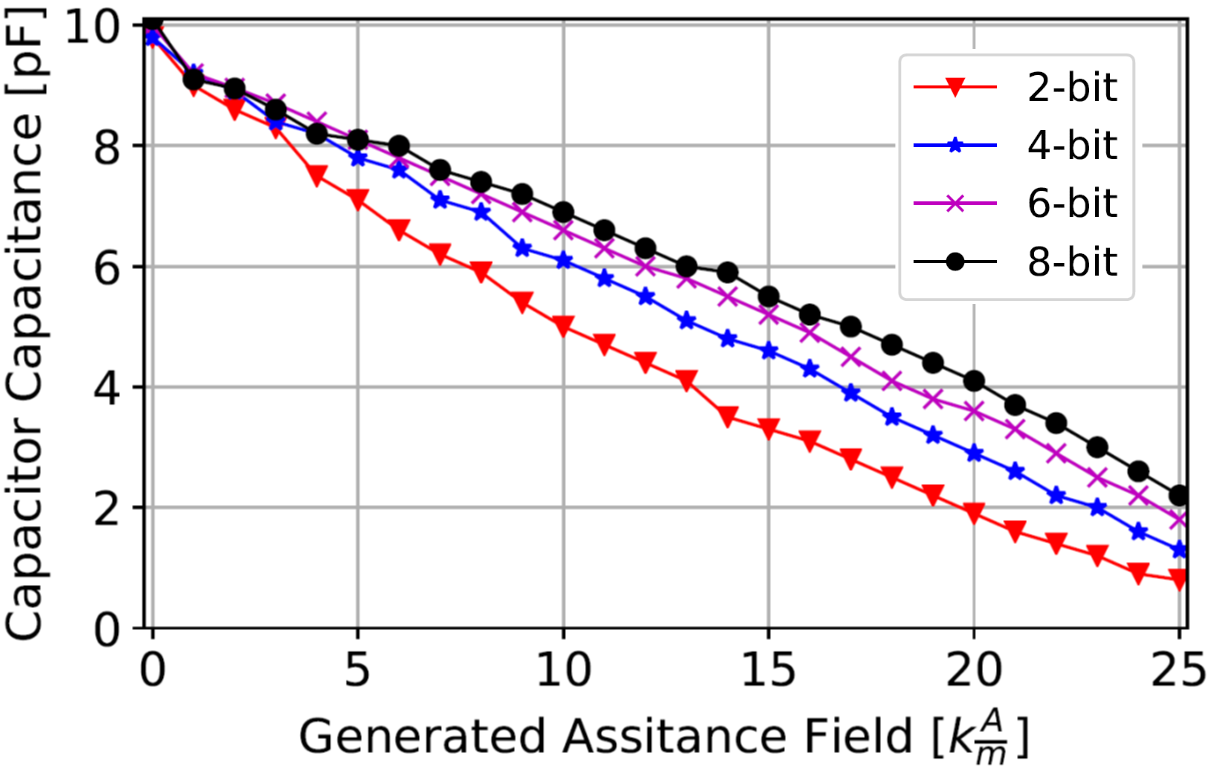}
\caption{The capacitor capacitance as a function of a generated assistance magnetic field while preserving the entropy. The field is in the direction of the fixed layer magnetization.}
\label{fig:c_vs_gen_field}
\vspace{-0.4cm}
\end{figure}

To reduce the area, the number of sense amplifiers can be lowered by reading the MTJs sequentially or by sharing sense amplifiers with a nearby MTJ memory array. However, the major area contributor is the capacitor $C$. Using a capacitor other than a MOS capacitor to reduce the area is left as future work. Alternatively, it is possible to lower the capacitance (and hence the size of the capacitor) while maintaining the same switching distribution, but this requires reducing the number of MTJs and\textbackslash or using larger transistors (reducing their resistance) to preserve the same current through the MTJs.
Another option is to use MTJ devices with a higher switching probability for the same current (as done in~\cite{Lee2017,Vodenicarevic}), which will require a smaller capacitor but will increase the sensitivity of the design to external magnetic fields. 
A different option is to use a generated magnetic field by a dedicated wire, similarly to the solution suggested in~\cite{Patel} for STT-MRAM memory. The magnetic field will raise the switching probability; hence, a smaller current will be needed to produce the same entropy, also resulting in a smaller capacitor. Figure~\ref{fig:c_vs_gen_field} shows the relationship between the generated magnetic field (in the direction of the fixed layer magnetization) and the capacitor required to achieve high entropy per bit (close to 1). Additionally, the generated magnetic field might be dynamically adjusted to compensate for the effects of process variation and parameter drift of the circuit. However, such a solution might incur an area overhead and design complications. Precise analysis of such a solution is left for future research.

The energy of the TRNG in the Reset and Read steps is for capacitor charging, switching from P to AP of the MTJs (at the Reset step), and for the read operation.
The Enable step only uses the energy stored in the capacitor. The energy required to charge the capacitor is the energy the capacitor holds, $3.2pJ$, plus another $3.2pJ$ consumed on the passgate connected to it (transistors $N1$ and $P1$ in Figure~\ref{pic:TRNG_circuit}), for a passgate effective resistance of $1.5K\Omega$ and Reset step time of $66ns$.
The MTJs have a write energy of $4.5pJ$ and a read energy of $0.7pJ$~\cite{Dong}. Table~\ref{table:cmpr} lists the energy per bit and power for different $N$.

\subsection{Comparison to Other TRNGs}
Table~\ref{table:cmpr} compares our proposed design with two different state-of-the-art CMOS based TRNGs. 
The proposed TRNG has a high entropy generation rate and low energy per bit as compared to CMOS TRNGs with a similar area and power.

The proposed TRNG is not compared with previously proposed STT-MTJ based TRNGs since they used specially designed STT-MTJ devices or did not consider process variation effects or external magnetic field effects (as discussed in Section~\ref{sec:related_work}).

Yang \textit{et al.}'s~\cite{KYang} TRNG is based on ring oscillators with a 28nm CMOS process node. The design of the TRNG includes several ROs, each controlling a 14-bit counter to measure an event (private for each RO). The resulting counter is random due to the noise in the RO and serves as the output. Only a subset of the counter's bits is used to produce a uniform distribution. This TRNG has similar area and power as our proposed TRNG.
The proposed 8-bit TRNG has $4\times$ to $5\times$ higher entropy rate, and  $3\times$ to $4\times$ better energy per bit than the TRNG of Yang \textit{et al.}~\cite{KYang}.

Srinivasan \textit{et al.}'s~\cite{IntelTRNG} TRNG is based on a metastable latch with a 45nm CMOS process node. It has $10\times$ larger area and consumes $10\times$ more power.
The metastable latch TRNG is $20\times-24\times$ faster and consumes $2\times-2.6\times$ less energy per bit than our 8-bit proposed design.
Nevertheless, the design proposed by Srinivasan \textit{et al.}~\cite{IntelTRNG} generates a single random bit every clock cycle, requiring the use of a $2.4GHz$ clock to achieve the high entropy rate. The use of a high rate clock and the relatively large area make this design impractical for low-power and low-frequency devices.  

\section{Dealing with an Adversary}
\label{sec:adversary}

Consider an attack model where the attacker can change the environmental conditions of the TRNG.
For example, the attacker can place a fixed magnet in proximity to the TRNG  to control the external magnetic field, or use an antenna, or remotely control a circuit close to the TRNG (such as a processor). However, we assume that the adversary does not have physically invasive access.

\begin{table}[t]
\centering
\caption{Comparison of entropy generation rate, area, energy, and power of the proposed TRNG\protect\footnote{The numbers for the proposed TRNG are presented for 90\% of instances and for the highest generating rate.} with CMOS TRNGs}
\vspace{-0.2cm}
\begin{tabular}{|@{\,}c@{\,}|c|c|c|c|} 
\hline
\rule{0pt}{2ex} 	  & \makecell{Entropy Generation\\Rate $[Mb/s]$}&\makecell{Area\\$[\mu m^2]$} & \makecell{Energy per bit\\$[\frac{pJ}{entropy\shortminus bit}]$}&\makecell{Power\\$[mW]$}\\ \hline\hline
\rule{0pt}{2ex} 2-bit TRNG	                                    & 15.4-33.4	    & 401.2 	& 6.16-13.4     & 0.2\\ \hline
\rule{0pt}{2ex} 4-bit TRNG	                                    & 40.9-63.8	    & 402.4 	& 5.9-9.2       & 0.38\\\hline
\rule{0pt}{2ex} 6-bit TRNG	                                    & 66.7-92.7	    & 403.6	    & 5.8-8.0       & 0.54\\\hline
\rule{0pt}{2ex} 8-bit TRNG	                                    & 94.0-120.6	& 404.8	    & 5.7-7.3       &0.7 \\\hline
\rule{0pt}{2ex} \makecell{Yang \textit{et al.} \\\cite{KYang}}  & 23.16 	    & 375	    & 23            &0.54\\ \hline
\rule{0pt}{2ex} \makecell{Srinivasan \\\textit{et al.} \cite{IntelTRNG}} & 2400 & 4004	    & 2.9           &7\\ \hline
\end{tabular}
\vspace{-0.4cm}
\label{table:cmpr}
\end{table}

Temperature has little effect on the TRNG (Section~\ref{subsec:eval}) and therefore is not an interesting attack venue. On the other hand, the external magnetic field on the TRNG can decrease the entropy substantially. Passive shielding can mitigate the effect of an external field. Prior work~\cite{Yamada,Wang,Paperno} on passive shielding demonstrated this for MTJ-based memories.

A different approach to interference is detection. This can be done using
online tests that check for a certain amount of randomness. Once the randomness is below a specific threshold, an error should be sent to the operating entity, informing it of nearby interference or an ongoing attack. This solution will not prevent the attack, but it might convert it to a denial-of-service attack.

\section{Conclusions}
In this paper, we presented an asynchronous TRNG that utilizes the random switching time of STT-MTJ devices. The TRNG was comprehensively evaluated in simulations using the physical equations describing the STT-MTJs. The evaluation showed that by increasing the number of STT-MTJs in the design, the TRNG can have greater entropy per output and better resilience to process variation. Furthermore, the design achieves better throughput than current CMOS TRNGs, with lower energy per bit and similar die area and power dissipation. However, MTJ devices are susceptible to attacks controlling the external magnetic field, requiring the use of additional countermeasures.  

\bibliographystyle{IEEEtran}
\bibliography{IEEEabrv,bibtex}

\vskip -1\baselineskip plus -1fil
\begin{IEEEbiography}[{\includegraphics[width=1in,height=1.25in,clip,keepaspectratio]{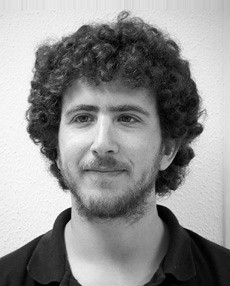}}]{Ben Perach}
received the B.Sc. degree in mathematics from The Hebrew University of Jerusalem in 2010 and the M.Sc. degree in electrical engineering from Tel Aviv University in 2017. He is currently working toward his Ph.D. degree in electrical engineering at the Technion – Israel Institute of Technology, Haifa, Israel. His current research interests include computer architecture with a focus on processor design, and also field-programmable gate arrays, security, and data networks.
\end{IEEEbiography}

\vskip -1\baselineskip plus -1fil

\begin{IEEEbiography}[{\includegraphics[width=1in,height=1.25in,clip,keepaspectratio]{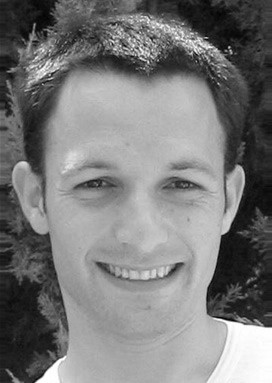}}]{Shahar Kvatinsky}
is an assistant professor at the Andrew and Erna Viterbi Faculty of Electrical Engineering, Technion – Israel Institute of Technology. He received the B.Sc. degree in computer engineering and applied physics and an MBA degree in 2009 and 2010, respectively, both from the Hebrew University of Jerusalem, and the Ph.D. degree in electrical engineering from the Technion – Israel Institute of Technology in 2014. From 2006 to 2009 he was with Intel as a circuit designer and was a post-doctoral research fellow at Stanford University from 2014 to 2015. Kvatinsky is an editor of Microelectronics Journal and has been the recipient of the 2015 IEEE Guillemin-Cauer Best Paper Award, 2015 Best Paper of Computer Architecture Letters, Viterbi Fellowship, Jacobs Fellowship, ERC starting grant, the 2017 Pazy Memorial Award, the 2014 and 2017 Hershel Rich Technion Innovation Awards, 2013 Sanford Kaplan Prize for Creative Management in High Tech, 2010 Benin prize, and seven Technion excellence teaching awards. His current research is focused on circuits and architectures with emerging memory technologies and design of energy efficient architectures.
\end{IEEEbiography}

\end{document}